\documentclass[a4paper,graphics,natbib,graphicx,psfig,amssymb,ccaption]{article}
\usepackage{lscape}
\usepackage{graphicx}
\newcommand{\affil}[1]{$^{\rm #1}$}
\renewcommand{\baselinestretch}{1.2}
\usepackage[authoryear]{natbib}
\bibpunct{(}{)}{;}{a}{}{,}
\date{} 

\title{\large\bf\flushleft Absolute Magnitude Calibration for Giants based on the Colour-Magnitude Diagrams of Galactic Clusters. II-Calibration with SDSS}
\author{\parbox{\textwidth}{\flushleft
\vspace{-0.5cm}
{\it S. Karaali\affil{\dag, A, B}, S. Bilir\affil{A}, and E. Yaz G\"ok\c ce\affil{A}}\\
\vspace{0.4cm}
{\small \affil{A}\,Istanbul University, Faculty of Sciences, Department of Astronomy and Space Sciences, 34119, Istanbul, Turkey}\\
{\small \affil{B}\,Email: karsa@istanbul.edu.tr}}}
\begin{document}
\twocolumn[
\begin{changemargin}{.8cm}{.5cm}
\begin{minipage}{.9\textwidth}
\vspace{-1cm}
\maketitle
\small{\bf Abstract:}
We present an absolute magnitude calibration for red giants with the colour magnitude diagrams of six Galactic clusters 
with different metallicities i.e. M92, M13, M3, M71, NGC 6791 and NGC 2158. The combination of the absolute magnitudes 
of the red giant sequences with the corresponding metallicities provides calibration for absolute magnitude estimation 
for red giants for a given $(g-r)_{0}$ colour. The calibration  is defined in the colour interval 0.45 $\leq(g-r)_{0}\leq$ 
1.30 mag and it covers the metallicity interval $-2.15\leq \lbrack Fe/H \rbrack \leq$ +0.37 dex. The absolute magnitude 
residuals obtained by the application of the procedure to another set of Galactic clusters lie in the interval 
$-0.28< \Delta M \leq +0.43$ mag. However, the range of 94\% of the residuals is shorter, $-0.1<\Delta M \leq+0.4$ mag. 
The mean and the standard deviation of (all) residuals are 0.169 and 0.140 mag, respectively. The derived relations are 
applicable to stars older than 2 Gyr, the age of the youngest calibrating cluster.                

\medskip
\medskip
\end{minipage}
\end{changemargin}
]
\small
\let\thefootnote\relax\footnote{\small \affil{\dag}\,Retired.}
\section{Introduction}
The distance of an astronomical object plays an important role in deriving absolute magnitudes 
of stars and determining the three dimensional structure of the Milky Way galaxy. The distance to a star can 
be evaluated by trigonometric or photometric parallaxes. Trigonometric parallaxes are only 
available for nearby stars where {\em Hipparcos} \citep{ESA97} is the main supplier for 
the data. For stars at large distances, the use of photometric parallaxes is unavoidable. 
In other words, the study of the Galactic structure is strictly tied to precise determination 
of absolute magnitudes. 

Different methods can be used for absolute magnitude determination where most of them  are 
devoted to dwarfs. The method used in the Str\"omgren's $uvby-\beta$ \citep{NS91} and in 
the {\em UBV} \citep{Laird88} photometry depends on the absolute magnitude offset from a 
standard main-sequence. In recent years the derivation of absolute magnitudes has been 
carried out by means of colour-absolute magnitude diagrams of some specific clusters 
whose metal abundances are generally adopted as the mean metal abundance of a Galactic 
population, such as thin, thick discs and halo. The studies of \cite{Phleps00} and \cite{Chen01} 
can be given as examples. A slightly different approach is that of \citet{Siegel02} 
where two relations, one for stars with solar-like abundances and another one for 
metal-poor stars were derived between $M_{R}$ and the colour index $R-I$, where 
$M_{R}$ is the absolute magnitude in the $R$ filter of Johnson system. For a star of 
given metallicity and colour, absolute magnitude can be estimated by linear interpolation 
of {\em two} ridgelines and by means of {\em linear} extrapolation beyond the metal-poor ridgeline. 

The most recent procedure used for absolute magnitude determination consists of finding 
the most likely values of the stellar parameters, given the measured atmospheric ones, 
and the time spent by a star in each region of the H-R diagram. In practice, researchers 
select the subset of isochrones with $[M/H]\pm \Delta_{[M/H]}$, where $\Delta_{[M/H]}$ is 
the estimated error on the metallicity, for each set of derived $T_{eff}$, $\log g$ and $[M/H]$. 
Then a Gaussian weight is associated to each point of the selected isochrones, which depends on 
the measured atmospheric parameters and the considered errors. This criterion allows the 
algorithm to select only the points whose values are closed by the pipeline of the corresponding 
survey such as RAVE. For details of this procedure we cite the works of \cite{Breddels10} and 
\cite{Zwitter10}. This procedure is based on many parameters. Hence it provides absolute 
magnitudes with high accuracy. Also it can be applied to both dwarf and giant stars simultaneously.

In \cite{Karaali03}, we presented a procedure for the photometric parallax estimation of 
dwarf stars which depends on the absolute magnitude offset from the main-sequence of 
the Hyades cluster. \cite{Bilir08} obtained the absolute magnitude calibrations of 
the thin disc main-sequence stars in the optical ($M_{V}$) and in the near-infrared ($M_{J}$) bands 
using the recent reduced Hipparcos astrometric data \citep{Leeuwen07}. \cite{Bilir09} 
derived a new luminosity colour relation based on trigonometric parallaxes for the thin disc 
main-sequence stars with Sloan Digital Sky Survey (SDSS) photometry. In \cite{Karaali12}, 
we used a similar procedure for the absolute magnitude estimation of red giants by using 
the $V_{0}, (B-V)_{0}$ apparent magnitude-colour diagrams of Galactic clusters with 
different metallicities. Here, we will estimate absolute magnitudes for red giants with 
$g_{0}, (g-r)_{0}$ colour-magnitude diagrams. Thus, we will give a chance to the 
researchers who work with SDSS photometry \citep{Fukugita96} for a direct estimation of 
the absolute magnitudes of the red giants. The outline of the paper is as follows. 
We present the data in Section 2. The procedure used for calibration is given in Section 3, 
and Section 4 is devoted to 
summary and discussion.

\section{Data}

Six clusters with different metallicities, i.e. M92, M13, M3, M71, NGC 6791, and NGC 2158, 
were selected for our program. The $g'$ and $r'$ magnitudes  for the first five clusters 
were taken from \cite{Clem08}. Clem, Vanden Berg \& Stetson observed the clusters in the 
$u'g'r'i'z'$ passbands with the MegaCam wide-field imager on the Canada-France-Hawaii 
Telescope. Whereas the $g$ and $r$ magnitudes for the cluster NGC 2158 were provided by the 
observation of the cluster on instrumental {\em ugriz} passbands \citep{Smolinski11}.
The two sets of passbands are very similar, but not quite identical. We derived the following 
equations by the transformations of \cite{Rider04} and transformed the $g'$ and $g'-r'$ data of
 \cite{Clem08} to the $g$ and $g-r$ data. Thus, we obtained a homogeneous set 
of data for absolute magnitude calibration.   

\begin{eqnarray}
      g=g'+0.060\lbrack(g'-r')-0.53\rbrack \nonumber,\\
 g-r= 1.060(g'-r') - 0.035(r'-i')-0.024.                                     
\end{eqnarray}

The range of the metallicity of the clusters given in iron abundance is $-2.15 \leq \lbrack Fe/H \rbrack \leq + 0.37$ dex. 
The $(g-M_{g})_{0}$ true distance modulus, $E(B-V)$ colour excess, and $\lbrack Fe/H \rbrack$ 
iron abundance for M92, M13, M3, M71, NGC 6791 are taken from the authors given in second order of the 
reference list in Table 1, whereas the ones for NGC 2158 are those of \cite{Smolinski11}. The $g$ and 
$g –- r$ data are presented in Table 2. We adopted $R=A_{V}/E(B-V)=3.1$ to convert the colour excess 
to the extinction. Although different numerical values appeared in the literature for specific regions 
of our Galaxy, a single value is applicable everywhere. Then, we used the equations $A_{g}/A_{V}=$ 1.199 
and $A_{r}/A_{V}=$ 0.858, $A_{i}/A_{V}=$ 0.639 of \cite{Fan99} to evaluate the total extinctions in $A_{g}$,
$A_{r}$ and $A_{i}$. Then, the equation for the selective extinction in {\em SDSS} is $E(g-r)/ A_{V} =$ 0.341. 

The $u'g'r'i'z'$ magnitudes for the clusters in \cite{Clem08} were given in ridge-lines. We plotted the 
transformed $g_{0}, (g-r)_{0}$ sequences on a diagram for each cluster and identified the giants by means of their 
positions in the diagram. Whereas, the fiducial red giant sequence of the cluster NGC 2158 given in Table 2, supplied 
by binning the transformed $g$ magnitudes and $g-r$ colours of 54 red giants in \cite{Smolinski11}. We, then fitted 
the fiducial sequence of giants to high degree polynomials. A fourth degree polynomial was sufficient for the clusters 
NGC 6791 and NGC 2158, whereas a fifth degree polynomial was necessary for a good correlation coefficient for the clusters 
M92, M13, M3, M71, for a good correlation coefficient. The calibration of $g_{0}$ is as follows:

\begin{table}
\setlength{\tabcolsep}{2pt}
\center
 \caption{Data for the clusters used in our work.}
\scriptsize{
 \label{tabledata}
  \begin{tabular}{lcccr}
    \hline
    Cluster & $E(B-V)$  & $(g-M_{g})_{0}$ & $\lbrack Fe/H \rbrack$ & Ref.\\
            & (mag)     & (mag)              & (dex)                  &  \\
    \hline
    M92     & 0.025 & 14.72 & -2.15 & (1), (2) \\
    M13     & 0.020 & 14.38 & -1.41 & (1), (2) \\
    M3      & 0.010 & 15.04 & -1.50 & (1), (3) \\
    M71     & 0.280 & 12.83 & -0.78 & (1), (4) \\
    NGC6791 & 0.100 & 12.94 & 0.37  & (1), (5) \\
    NGC2158 & 0.440 & 12.80 & -0.25 &      (6) \\
    \hline
\end{tabular}\\
(1) \cite{Clem08}, (2) \cite{Gratton97}, (3) \cite{Harris96,Harris10}, (4) \cite{Hodder92}, (5) \cite{Sandage03}, (6) \cite{Smolinski11}.
\label{tab:addlabel}
}
\end{table}

\begin{eqnarray}
   g_{0}= \sum_{i=0}^{5}a_{i}(g-r)^{i}_0. 
\end{eqnarray}

The numerical values of the coefficients $a_{i}$ ($i=$ 0, 1, 2, 3, 4, 5) are given in Table 3 and the corresponding diagrams are presented in Fig. 1. The ($g-r)_{0}$-interval in the second line of the table denotes the range of ($g-r)_{0}$ available for each cluster. 

\begin{table*}
  \center
\setlength{\tabcolsep}{1.5pt}
  \caption{Original $g'$, $g'-r'$, $r'-i'$ and the transformed $g_{0}$, $(g-r)_{0}$ data  for the clusters M92, M13, M3, M71 and NGC 6791. The $g$ magnitudes and $g-r$ colours for the cluster NGC 2158 are original.}

\scriptsize{
    \begin{tabular}{cccccccc|cccccccc}
    \hline
$g'-r'$&$g'$& $r'-i'$&$g'_{0}$&$(g'-r')_{0}$&$(r'-i')_{0}$&$(g-r)_{0}$&$g_{0}$&$g'-r'$&$g'$&$r'-i'$&$g'_{0}$&$(g'-r')_{0}$&$(r'-i')_{0}$&$(g-r)_{0}$&$g_{0}$\\
    \hline
    \multicolumn{8}{c|}{M92}                                      & \multicolumn{8}{c}{M3 (cont.)}                                     \\
    \hline
    1.150 & 12.650 & 0.508 & 12.557 & 1.124 & 0.491 & 1.150 & 12.593 & 0.487 & 17.987 & 0.192 & 17.950 & 0.476 & 0.185 & 0.475 & 17.947 \\
    0.952 & 12.952 & 0.411 & 12.859 & 0.926 & 0.394 & 0.943 & 12.883 & 0.479 & 18.179 & 0.189 & 18.142 & 0.468 & 0.182 & 0.466 & 18.138 \\
    0.835 & 13.335 & 0.361 & 13.242 & 0.809 & 0.344 & 0.821 & 13.259 & 0.467 & 18.367 & 0.184 & 18.330 & 0.456 & 0.177 & 0.454 & 18.325 \\
\cline{9-16}    0.746 & 13.746 & 0.330 & 13.653 & 0.720 & 0.313 & 0.728 & 13.664 & \multicolumn{8}{c}{M71}                              \\
\cline{9-16}    0.684 & 14.184 & 0.300 & 14.091 & 0.658 & 0.283 & 0.663 & 14.099 & 1.547 & 13.047 & 0.854 & 12.006 & 1.251 & 0.664 & 1.279 & 12.050\\
    0.634 & 14.634 & 0.276 & 14.541 & 0.608 & 0.259 & 0.611 & 14.546 & 1.344 & 13.344 & 0.640 & 12.303 & 1.048 & 0.450 & 1.071 & 12.334 \\
    0.593 & 15.093 & 0.255 & 15.000 & 0.567 & 0.238 & 0.568 & 15.002 & 1.203 & 13.703 & 0.566 & 12.662 & 0.907 & 0.376 & 0.924 & 12.685 \\
    0.554 & 15.554 & 0.238 & 15.461 & 0.528 & 0.221 & 0.527 & 15.461 & 1.102 & 14.102 & 0.511 & 13.061 & 0.806 & 0.321 & 0.819 & 13.078 \\
    0.523 & 16.023 & 0.222 & 15.930 & 0.497 & 0.205 & 0.495 & 15.928 & 1.019 & 14.519 & 0.464 & 13.478 & 0.723 & 0.274 & 0.733 & 13.490 \\
    0.497 & 16.497 & 0.210 & 16.404 & 0.471 & 0.193 & 0.468 & 16.401 & 0.953 & 14.953 & 0.439 & 13.912 & 0.657 & 0.249 & 0.664 & 13.920 \\
    0.476 & 16.976 & 0.200 & 16.883 & 0.450 & 0.183 & 0.446 & 16.878 & 0.907 & 15.407 & 0.423 & 14.366 & 0.611 & 0.233 & 0.616 & 14.371 \\
    0.458 & 17.458 & 0.189 & 17.365 & 0.432 & 0.172 & 0.427 & 17.359 & 0.871 & 15.871 & 0.412 & 14.830 & 0.575 & 0.222 & 0.578 & 14.833 \\
    0.448 & 17.648 & 0.184 & 17.555 & 0.422 & 0.167 & 0.417 & 17.549 & 0.837 & 16.337 & 0.405 & 15.296 & 0.541 & 0.215 & 0.542 & 15.297 \\
\cline{1-8}    \multicolumn{8}{c|}{M13}                              & 0.813 & 16.813 & 0.399 & 15.772 & 0.517 & 0.209 & 0.517 & 15.771 \\
\cline{1-8}    1.330 & 12.630 & 0.540 & 12.556 & 1.309 & 0.526 & 1.345 & 12.602 & 0.795 & 17.295 & 0.394 & 16.254 & 0.499 & 0.204 & 0.498 & 16.252 \\
    1.071 & 12.871 & 0.460 & 12.797 & 1.050 & 0.446 & 1.073 & 12.828 & 0.789 & 17.489 & 0.391 & 16.448 & 0.493 & 0.201 & 0.492 & 16.446 \\
    0.933 & 13.233 & 0.398 & 13.159 & 0.912 & 0.384 & 0.929 & 13.182 & 0.781 & 17.681 & 0.381 & 16.640 & 0.485 & 0.191 & 0.483 & 16.638 \\
\cline{9-16}    0.824 & 13.624 & 0.353 & 13.550 & 0.803 & 0.339 & 0.815 & 13.566 & \multicolumn{8}{c}{NGC 6791}                          \\
\cline{9-16}    0.750 & 14.050 & 0.320 & 13.976 & 0.729 & 0.306 & 0.738 & 13.988 & 1.414 & 14.514 & 0.730 & 14.142 & 1.308 & 0.662 & 1.340 & 14.189 \\
    0.694 & 14.494 & 0.292 & 14.420 & 0.673 & 0.278 & 0.679 & 14.428 & 1.298 & 14.898 & 0.543 & 14.526 & 1.192 & 0.475 & 1.223 & 14.566 \\
    0.647 & 14.947 & 0.268 & 14.873 & 0.626 & 0.254 & 0.631 & 14.878 & 1.214 & 15.314 & 0.468 & 14.942 & 1.108 & 0.400 & 1.137 & 14.977 \\
    0.607 & 15.407 & 0.250 & 15.333 & 0.586 & 0.236 & 0.589 & 15.336 & 1.145 & 15.745 & 0.424 & 15.373 & 1.039 & 0.356 & 1.065 & 15.404 \\
    0.570 & 15.870 & 0.235 & 15.796 & 0.549 & 0.221 & 0.550 & 15.797 & 1.086 & 16.186 & 0.389 & 15.814 & 0.980 & 0.321 & 1.004 & 15.841 \\
    0.538 & 16.338 & 0.224 & 16.264 & 0.517 & 0.210 & 0.517 & 16.263 & 1.040 & 16.640 & 0.363 & 16.268 & 0.934 & 0.295 & 0.956 & 16.293 \\
    0.509 & 16.809 & 0.214 & 16.735 & 0.488 & 0.200 & 0.486 & 16.732 & 1.006 & 17.106 & 0.348 & 16.734 & 0.900 & 0.280 & 0.921 & 16.757 \\
    0.486 & 17.286 & 0.206 & 17.212 & 0.465 & 0.192 & 0.462 & 17.208 & 0.977 & 17.577 & 0.341 & 17.205 & 0.871 & 0.273 & 0.890 & 17.226 \\
    0.474 & 17.574 & 0.200 & 17.500 & 0.453 & 0.186 & 0.450 & 17.495 & 0.965 & 17.765 & 0.339 & 17.393 & 0.859 & 0.271 & 0.877 & 17.413 \\
    0.461 & 17.761 & 0.195 & 17.687 & 0.440 & 0.181 & 0.436 & 17.681 & 0.946 & 17.946 & 0.335 & 17.574 & 0.840 & 0.267 & 0.857 & 17.593 \\
    \hline
    \multicolumn{8}{c|}{M3}                                       & \multicolumn{8}{c}{NGC 2158}                                         \\
    \hline
    1.300 & 13.300 & 0.580 & 13.263 & 1.289 & 0.573 & 1.323 & 13.308 & $g-r$   & $g$ & $(g-r)_{0}$&$g_{0}$   &    &    &     &         \\
\cline{9-16}    1.058 & 13.558 & 0.446 & 13.521 & 1.047 & 0.439 & 1.071 & 13.552 & 0.553 & 15.739 & 0.553 & 15.739 & $-$    & $-$    & $-$    & $-$ \\
    0.918 & 13.918 & 0.383 & 13.881 & 0.907 & 0.376 & 0.925 & 13.903 & 0.614 & 15.294 & 0.614 & 15.294 & $-$    & $-$    & $-$    & $-$ \\
    0.810 & 14.310 & 0.338 & 14.273 & 0.799 & 0.331 & 0.812 & 14.289 & 0.708 & 14.777 & 0.708 & 14.777 & $-$    & $-$    & $-$    & $-$ \\
    0.733 & 14.733 & 0.305 & 14.696 & 0.722 & 0.298 & 0.731 & 14.707 & 0.741 & 14.508 & 0.741 & 14.508 & $-$    & $-$    & $-$    & $-$ \\
    0.675 & 15.175 & 0.281 & 15.138 & 0.664 & 0.274 & 0.671 & 15.146 & 0.763 & 14.345 & 0.783 & 14.297 & $-$    & $-$    & $-$    & $-$ \\
    0.626 & 15.626 & 0.261 & 15.589 & 0.615 & 0.254 & 0.619 & 15.594 & 0.782 & 14.265 & 0.852 & 13.870 & $-$    & $-$    & $-$    & $-$ \\
    0.584 & 16.084 & 0.245 & 16.047 & 0.573 & 0.238 & 0.575 & 16.049 & 0.803 & 14.296 & 0.965 & 13.687 & $-$    & $-$    & $-$    & $-$ \\
    0.555 & 16.555 & 0.230 & 16.518 & 0.544 & 0.223 & 0.545 & 16.519 & 0.852 & 13.870 & $-$   & $-$    & $-$    & $-$    & $-$    & $-$ \\
    0.528 & 17.028 & 0.215 & 16.991 & 0.517 & 0.208 & 0.517 & 16.990 & 0.965 & 13.687 & $-$   & $-$    & $-$    & $-$    & $-$    & $-$ \\
    0.507 & 17.507 & 0.203 & 17.470 & 0.496 & 0.196 & 0.495 & 17.468 & $-$    & $-$   & $-$   & $-$    & $-$    & $-$    & $-$    & $-$ \\
    \hline
    \end{tabular}
}
  \label{cl-calibration}
\end{table*}

\begin{figure}
\begin{center}
\includegraphics[scale=0.35, angle=0]{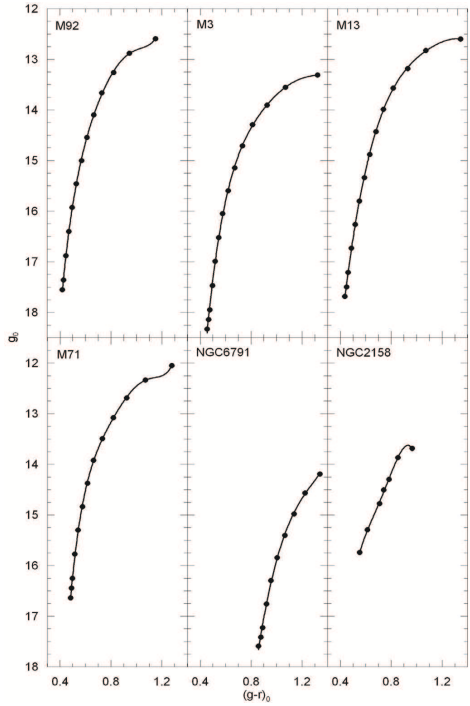} 
\caption[] {$g_{0}, (g-r)_{0}$ colour-apparent magnitude diagrams for six Galactic clusters used for the absolute magnitude calibration.} 
\label{his:col}
\end{center}
\end {figure}

\begin{table}[h]
\setlength{\tabcolsep}{2.5pt}
  \center
\tiny{
  \caption{Numerical values of the coefficients $a_{i}$ ($i=$0, 1, 2, 3, 4, 5).}
    \begin{tabular}{lrrrrrr}
    \hline
 Cluster & M92   & M13   & M3    & M71   & NGC 6791 & NGC 2158\\
    \hline
$(g-r)_{0}$ & [0.42-1.15] & [0.44-1.35] & [0.45-1.32] & [0.48-1.28] & [0.86-1.34] & [0.55-0.96]\\
     interval &           &             &             &             &             &            \\  
  \hline
    
  $a_{5}$   & -107.410 &   1.2115 &  -30.463 &  -133.31 &    $-$   &      $-$ \\
  $a_{4}$   &  442.770 &   5.9697 &  161.410 &   617.40 &  -3.2426 &  201.770 \\
  $a_{3}$   & -729.420 & -41.8770 & -340.950 & -1132.00 &  -7.3559 & -581.310 \\
  $a_{2}$   &  609.900 &  81.6260 &  362.750 &  1033.00 &  60.9390 &  624.230 \\
  $a_{1}$   & -266.560 & -69.8160 & -198.370 &  -475.38 & -96.0120 & -302.130 \\
  $a_{0}$   &   63.533 &  35.8660 &   59.319 &   102.71 &  61.5710 &   71.356 \\
  \hline
    \end{tabular}
}
  \label{tab:addlabel}
\end{table}

\begin{figure}
\begin{center}
\includegraphics[scale=0.35, angle=0]{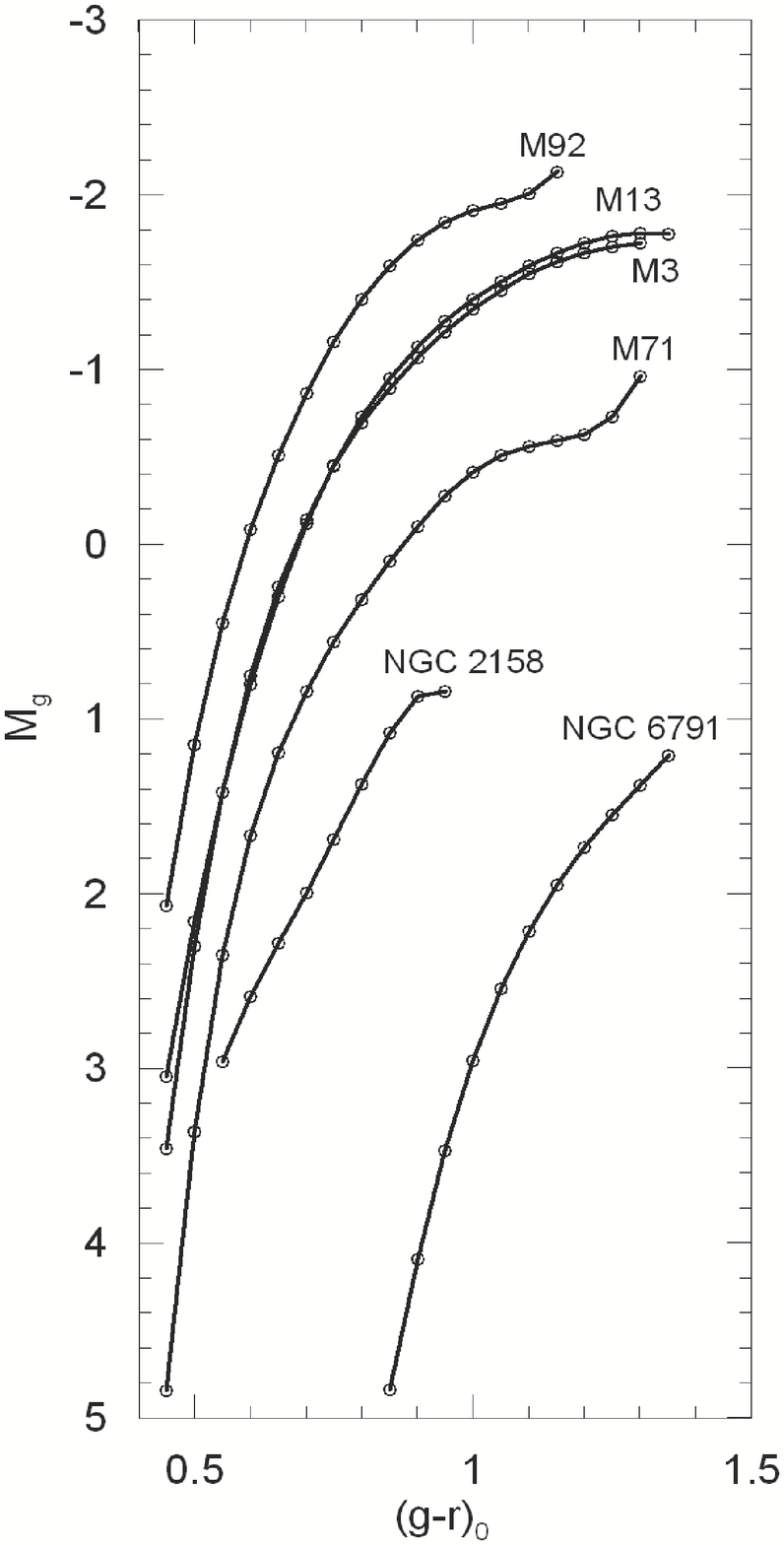} 
\caption[] {$M_{g}, (g-r)_{0}$ colour-absolute magnitude diagrams for six clusters used for the absolute magnitude calibration.} 
\label{his:col}
\end{center}
\end {figure}

\section{The Procedure}
\subsection{Absolute Magnitude as a Function of Metallicity}

The procedure consists of a slight modification of the procedure in Paper I. There, we calibrated 
the absolute magnitude offsets from the fiducial red giant sequence of a standard cluster (M5) for 
a given colour index as a function of metallicity offsets. Whereas, here we calibrated the absolute 
magnitudes directly to metallicities for a given $(g-r)_{0}$ colour. Thus, one does not need 
to calculate an absolute magnitude offset from a standard cluster and then to add it to the 
corresponding absolute magnitude of the standard cluster for the final absolute magnitude 
estimation. Also, the new procedure decreased the number of columns in the final tables. We 
estimated the $M_{g}$ absolute magnitudes for the $(g-r)_{0}$ colours 
given in Table 4 for the cluster sample in Table 1 by combining the $g_{0}$ apparent magnitudes 
evaluated by Eq. (2) and the true distance modulus ($\mu_{0}$) of the cluster in question, i.e.

\begin{eqnarray}
M_{g}=g_{0}-\mu_{0}.     
\end{eqnarray}
Then, we plotted the absolute magnitudes versus $(g-r)_{0}$ colours. Fig. 2 shows that the absolute magnitude 
is colour and metallicity dependent. It increases (algebraically) with increasing metallicity and decreasing colour.

Now, we can fit the $M_{g}$ absolute magnitudes to the corresponding $\lbrack Fe/H \rbrack$ metallicity for 
a given $(g-r)_{0}$ colour index and obtain the required calibration. This is carried out for the colour 
indices $(g-r)_{0}=$ 0.60, 0.75, 0.95, 1.05 and 1.20 just for the exhibition of the procedure. The results 
are given in Table 5 and Fig. 3. The absolute magnitudes in the colour indices $(g-r)_{0}=$ 0.60, 0.75, 1.05 
and 1.20 could be fitted to a second degree polynomial with (squared) correlation coefficients $R^{2}\geq $ 0.9993.  
The range of the metallicity for the colour index $(g-r)_{0}=$ 0.95 is the highest, i.e. 
$-2.15 \leq \lbrack Fe/H \rbrack \leq 0.37$ dex. Hence, a third degree polynomial was necessary for the high 
(squared) correlation coefficient $R^{2}=$ 0.9994. The high correlation coefficients indicate accurate 
absolute magnitude estimation. 

This procedure can be applied to any $(g-r)_{0}$ colour-interval for which the sample clusters are defined. 
The $(g-r)_{0}$ domain of the clusters are different. Hence, we adopted this interval in our study as 
0.45 $ \leq (g-r)_{0} \leq$ 1.30 where at least two clusters are defined, and we evaluated $M_{g}$ absolute 
magnitudes for each colour. Then, we combined them with the corresponding $\lbrack Fe/H \rbrack$ metallicities 
and obtained the final calibrations. The metallicities of the clusters M13 and M3 are close to each other, i.e. 
$\lbrack Fe/H \rbrack$ -1.41 and -1.50 dex, respectively. Hence, we adopted the mean of the data of these clusters 
in the absolute magnitude calibration. The general form of the equation for the calibrations is as follows:

\begin{eqnarray}
M_{g} = b_{0} + b_{1}X + b_{2}X^{2} + b_{3}X^{3}     
\end{eqnarray}
where $X = \lbrack Fe/H \rbrack$.         

$M_{g}$ could be fitted in terms metallicity by different degrees of polynomials. A cubic polynomial was necessary  
only for a limited interval, i.e. 0.85 $\leq (g-r)_{0} \leq$ 0.96, for a high correlation coefficient. Whereas a 
quadratic or linear polynomials were sufficient for most of the colour indices, i.e. 0.45 $\leq (g-r)_{0} \leq$ 0.84 
and 0.97 $\leq (g-r)_{0} \leq$ 1.30, for a high correlation coefficient. The degree of the polynomial depends mainly 
on the metallicity range considered. However, despite of the large domain in metallicity, 
-2.15 $\leq \lbrack Fe/H \rbrack \leq$ 0.37 dex, absolute magnitudes for the colour index interval 
0.97 $\leq (g-r)_{0} \leq$ 1.15 could be fitted by quadratic polynomials with high correlation coefficients. Then, 
one can say that the data presented for different clusters are homogeneous and our procedure promises accurate 
absolute magnitude estimation. The absolute magnitudes estimated via Eq. (3) for 86 $(g-r)_{0}$ colour indices and 
the corresponding $b_{i}$ ($i=$0, 1, 2, 3) coefficients are given in Table 6. However, the diagrams for the 
calibrations are not given in the paper for avoiding space consuming. One can use any data set taken from Table 6 
depending on the desire accuracy, and apply it to stars whose iron abundances are available. 

The calibration of $M_{g}$ in terms of $\lbrack Fe/H \rbrack$ is carried out for the colour interval 
0.45 $\leq (g-r)_{0} \leq$ 1.30 mag in steps of 0.01 mag. A small step is necessary to isolate 
an observational error on $g-r$ plus an error due to reddening. The origin of the mentioned errors is 
the trend of the red giant branch (RGB) sequence. As it is very steep, a small error in $g-r$ implies 
a large change in the absolute magnitude.

Iron abundance, $\lbrack Fe/H \rbrack$, is not the only parameter determining the chemistry of the star but also alpha 
enhancement, $\lbrack \alpha /Fe \rbrack$, is surely important. However, as stated in Paper I, there is a correlation 
between two sets of abundances. Hence, we do not expect any considerable change in the numerical values of $M_{g}$ 
in the case of addition of the alpha enhancement term in Eq. (4).

\begin{table*}
\setlength{\tabcolsep}{5pt}
  \center
  \caption{$M_{g}$ absolute magnitudes estimated for a set of $(g-r)_{0}$ colours for six Galactic clusters used in the calibration.}
    \begin{tabular}{ccccccc}
    \hline
    Cluster $\rightarrow$ & M92   & M13   & M3    & M71   & NGC 2158 & NGC 6791 \\
    \hline
    $(g-r)_{0}$ &  \multicolumn{6}{c}{$M_{g}$}               \\
    \hline
    0.45  &  2.072 &  3.049 &  3.457 &  4.845 & $-$    & $-$ \\
    0.50  &  1.147 &  2.161 &  2.299 &  3.362 & $-$    & $-$ \\
    0.55  &  0.453 &  1.419 &  1.419 &  2.354 & 2.962  & $-$ \\
    0.60  & -0.083 &  0.804 &  0.752 &  1.669 & 2.587  & $-$\\
    0.65  & -0.511 &  0.298 &  0.245 &  1.192 & 2.284  & $-$ \\
    0.70  & -0.862 & -0.115 & -0.144 &  0.840 & 1.993  & $-$ \\
    0.75  & -1.156 & -0.452 & -0.448 &  0.559 & 1.689  & $-$ \\
    0.80  & -1.400 & -0.725 & -0.692 &  0.316 & 1.373  & $-$ \\
    0.85  & -1.595 & -0.947 & -0.894 &  0.096 & 1.080  & 4.839 \\
    0.90  & -1.742 & -1.128 & -1.066 & -0.102 & 0.872  & 4.091 \\
    0.95  & -1.844 & -1.276 & -1.215 & -0.274 & 0.842  & 3.469 \\
    1.00  & -1.907 & -1.400 & -1.344 & -0.410 & $-$    & 2.960 \\
    1.05  & -1.951 & -1.504 & -1.454 & -0.506 & $-$    & 2.547 \\
    1.10  & -2.007 & -1.591 & -1.546 & -0.562 & $-$    & 2.216 \\
    1.15  & -2.127 & -1.664 & -1.617 & -0.593 & $-$    & 1.950 \\
    1.20  &  $-$   & -1.722 & -1.669 & -0.629 & $-$    & 1.734 \\
    1.25  &  $-$   & -1.763 & -1.703 & -0.726 & $-$    & 1.550 \\
    1.30  &  $-$   & -1.782 & -1.726 & -0.963 & $-$    & 1.380 \\
    1.35  &  $-$   & -1.775 &  $-$   &  $-$   & $-$    & 1.208 \\
    \hline

    \end{tabular}
  \label{tab:addlabel}
\end{table*}

\begin{table}
  \center
  \caption{$M_{g}$ absolute magnitudes and $\lbrack Fe/H \rbrack$ metallicities for five $(g-r)_{0}$-intervals }
    \begin{tabular}{lcc}
    \hline
    $(g-r)_{0}$ & $\lbrack Fe/H \rbrack$ & $M_{g}$\\
      (mag)    &  (dex)   &  (mag)  \\
    \hline
    0.60  & -2.15 & -0.083 \\
          & -1.46 &  0.778 \\
          & -0.78 &  1.669 \\
          & -0.25 &  2.587 \\
    \hline
    0.75  & -2.15 & -1.156 \\
          & -1.46 & -0.450 \\
          & -0.78 &  0.559 \\
          & -0.25 &  1.689 \\
    \hline
    0.95  & -2.15 & -1.844 \\
          & -1.46 & -1.246 \\
          & -0.78 & -0.274 \\
          & -0.25 &  0.842 \\
          & 0.37  &  3.469 \\
    \hline
    1.05  & -2.15 & -1.951 \\
          & -1.46 & -1.479 \\
          & -0.78 & -0.506 \\
          &  0.37 &  2.547 \\
    \hline
    1.20  & -1.46 & -1.695 \\
          & -0.78 & -0.629 \\
          &  0.37 &  1.734 \\
    \hline
    \end{tabular}
  \label{tab:addlabel}
\end{table}

\begin{table*}
\setlength{\tabcolsep}{1.5pt}
\renewcommand{\baselinestretch}{1}  
 \center
\tiny{
\caption{$M_{g}$ absolute magnitudes estimated for six Galactic clusters and the numerical values of $b_{i}$ ($i=0, 1, 2, 3$) coefficients in Eq. (3).
The absolute magnitudes and metallicities of the clusters M13 and M3 were combined in the evaluation of $b_{i}$ coefficients. The last column gives the range of the metallicity $[Fe/H]$ (dex) for the star whose absolute magnitude would be estimated. $R^{2}$ is the square of the correlation coefficient.} 
    \begin{tabular}{cccccccccccc}
    \hline
  Cluster $\rightarrow$  & M92   & M13+M3   & M71    & NGC 2158 & NGC 6791 &        &         &         &         &       &                 \\
    \hline
$(g-r)_{0}$ &            \multicolumn{5}{c}{$M_{g}$}        &$b_{0}$ & $b_{1}$ & $b_{2}$ & $b_{3}$ & $R^2$ & $[Fe/H]$-interval \\
    \hline
    0.45  &  2.072 &  3.253 &  4.845 & $-$   & $-$   & 7.2311 & 3.4339 & 0.4810 & $-$    & 1      & [-2.15, -0.78] \\
    0.46  &  1.864 &  3.029 &  4.502 & $-$   & $-$   & 6.6217 & 3.0054 & 0.3686 & $-$    & 1      & [-2.15, -0.78] \\
    0.47  &  1.669 &  2.815 &  4.183 & $-$   & $-$   & 6.0756 & 2.6404 & 0.2747 & $-$    & 1      & [-2.15, -0.78] \\
    0.48  &  1.484 &  2.611 &  3.888 & $-$   & $-$   & 5.2247 & 1.7538 & $-$    & $-$    & 0.9980 & [-2.15, -0.78] \\
    0.49  &  1.311 &  2.416 &  3.615 & $-$   & $-$   & 4.9045 & 1.6812 & $-$    & $-$    & 0.9990 & [-2.15, -0.78] \\
    0.50  &  1.147 &  2.230 &  3.362 & $-$   & $-$   & 4.6084 & 1.6161 & $-$    & $-$    & 0.9996 & [-2.15, -0.78] \\
    0.51  &  0.992 &  2.052 &  3.127 & $-$   & $-$   & 4.3349 & 1.5582 & $-$    & $-$    & 0.9998 & [-2.15, -0.78] \\
    0.52  &  0.846 &  1.883 &  2.911 & $-$   & $-$   & 4.0823 & 1.5068 & $-$    & $-$    & 1      & [-2.15, -0.78] \\
    0.53  &  0.708 &  1.721 &  2.710 & $-$   & $-$   & 3.8493 & 1.4615 & $-$    & $-$    & 1      & [-2.15, -0.78] \\
    0.54  &  0.577 &  1.566 &  2.525 & $-$   & $-$   & 3.6344 & 1.4219 & $-$    & $-$    & 1      & [-2.15, -0.78] \\
    0.55  &  0.453 &  1.419 &  2.354 & 2.962 & $-$   & 3.2590 & 1.1272 & -0.0842& $-$    & 0.9997 & [-2.15, -0.25] \\
    0.56  &  0.335 &  1.278 &  2.195 & 2.877 & $-$   & 3.2028 & 1.2831 & -0.0241& $-$    & 1      & [-2.15, -0.25] \\
    0.57  &  0.223 &  1.144 &  2.048 & 2.799 & $-$   & 3.1513 & 1.4305 & 0.0322 & $-$    & 1      & [-2.15, -0.25] \\
    0.58  &  0.116 &  1.016 &  1.912 & 2.724 & $-$   & 3.1036 & 1.5689 & 0.0846 & $-$    & 0.9998 & [-2.15, -0.25] \\
    0.59  &  0.014 &  0.894 &  1.786 & 2.654 & $-$   & 3.0586 & 1.6982 & 0.1330 & $-$    & 0.9996 & [-2.15, -0.25] \\
    0.60  & -0.083 &  0.778 &  1.669 & 2.587 & $-$   & 3.0158 & 1.8181 & 0.1774 & $-$    & 0.9993 & [-2.15, -0.25] \\
    0.61  & -0.176 &  0.667 &  1.560 & 2.523 & $-$   & 2.9743 & 1.9285 & 0.2180 & $-$    & 0.9991 & [-2.15, -0.25] \\
    0.62  & -0.265 &  0.561 &  1.458 & 2.461 & $-$   & 2.9335 & 2.0292 & 0.2547 & $-$    & 0.9989 & [-2.15, -0.25] \\
    0.63  & -0.350 &  0.460 &  1.364 & 2.401 & $-$   & 2.8927 & 2.1202 & 0.2875 & $-$    & 0.9988 & [-2.15, -0.25] \\
    0.64  & -0.432 &  0.364 &  1.275 & 2.342 & $-$   & 2.8515 & 2.2015 & 0.3167 & $-$    & 0.9987 & [-2.15, -0.25] \\
    0.65  & -0.511 &  0.272 &  1.192 & 2.284 & $-$   & 2.8093 & 2.2731 & 0.3421 & $-$    & 0.9987 & [-2.15, -0.25] \\
    0.66  & -0.586 &  0.184 &  1.114 & 2.226 & $-$   & 2.7658 & 2.3352 & 0.3640 & $-$    & 0.9988 & [-2.15, -0.25] \\
    0.67  & -0.659 &  0.100 &  1.040 & 2.168 & $-$   & 2.7206 & 2.3878 & 0.3824 & $-$    & 0.9988 & [-2.15, -0.25] \\
    0.68  & -0.729 &  0.020 &  0.970 & 2.110 & $-$   & 2.6735 & 2.4313 & 0.3975 & $-$    & 0.9989 & [-2.15, -0.25] \\
    0.69  & -0.797 & -0.056 &  0.904 & 2.052 & $-$   & 2.6242 & 2.4658 & 0.4094 & $-$    & 0.9991 & [-2.15, -0.25] \\
    0.70  & -0.862 & -0.130 &  0.840 & 1.993 & $-$   & 2.5726 & 2.4916 & 0.4183 & $-$    & 0.9992 & [-2.15, -0.25] \\
    0.71  & -0.925 & -0.199 &  0.780 & 1.934 & $-$   & 2.5187 & 2.5093 & 0.4244 & $-$    & 0.9993 & [-2.15, -0.25] \\
    0.72  & -0.986 & -0.266 &  0.722 & 1.874 & $-$   & 2.4623 & 2.5191 & 0.4278 & $-$    & 0.9994 & [-2.15, -0.25] \\
    0.73  & -1.045 & -0.330 &  0.666 & 1.813 & $-$   & 2.4037 & 2.5216 & 0.4287 & $-$    & 0.9996 & [-2.15, -0.25] \\
    0.74  & -1.101 & -0.391 &  0.612 & 1.751 & $-$   & 2.3428 & 2.5174 & 0.4274 & $-$    & 0.9997 & [-2.15, -0.25] \\
    0.75  & -1.156 & -0.450 &  0.559 & 1.689 & $-$   & 2.2799 & 2.5071 & 0.4242 & $-$    & 0.9998 & [-2.15, -0.25] \\
    0.76  & -1.209 & -0.506 &  0.508 & 1.626 & $-$   & 2.2152 & 2.4914 & 0.4192 & $-$    & 0.9998 & [-2.15, -0.25] \\
    0.77  & -1.259 & -0.560 &  0.458 & 1.563 & $-$   & 2.1491 & 2.4709 & 0.4128 & $-$    & 0.9999 & [-2.15, -0.25] \\
    0.78  & -1.308 & -0.611 &  0.410 & 1.500 & $-$   & 2.0819 & 2.4466 & 0.4052 & $-$    & 1      & [-2.15, -0.25] \\
    0.79  & -1.355 & -0.661 &  0.363 & 1.436 & $-$   & 2.0140 & 2.4193 & 0.3968 & $-$    & 1      & [-2.15, -0.25] \\
    0.80  & -1.400 & -0.708 &  0.316 & 1.373 & $-$   & 1.9460 & 2.3900 & 0.3880 & $-$    & 1      & [-2.15, -0.25] \\
    0.81  & -1.442 & -0.754 &  0.270 & 1.311 & $-$   & 1.8784 & 2.3595 & 0.3789 & $-$    & 1      & [-2.15, -0.25] \\
    0.82  & -1.484 & -0.798 &  0.226 & 1.251 & $-$   & 1.8119 & 2.3291 & 0.3701 & $-$    & 1      & [-2.15, -0.25] \\
    0.83  & -1.523 & -0.840 &  0.182 & 1.191 & $-$   & 1.7471 & 2.2999 & 0.3619 & $-$    & 1      & [-2.15, -0.25] \\
    0.84  & -1.560 & -0.881 &  0.139 & 1.134 & $-$   & 1.6848 & 2.2731 & 0.3547 & $-$    & 0.9999 & [-2.15, -0.25] \\
    0.85  & -1.595 & -0.920 &  0.096 & 1.080 & 4.839 & 2.3718 & 5.2610 & 3.2390 & 0.7703 & 0.9963 & [-2.15, 0.37] \\
    0.86  & -1.628 & -0.958 &  0.055 & 1.029 & 4.679 & 2.2861 & 5.1147 & 3.1103 & 0.7367 & 0.9964 & [-2.15, 0.37] \\
    0.87  & -1.660 & -0.995 &  0.014 & 0.982 & 4.524 & 2.2046 & 4.9722 & 2.9820 & 0.7027 & 0.9965 & [-2.15, 0.37] \\
    0.88  & -1.689 & -1.030 & -0.026 & 0.939 & 4.374 & 2.1278 & 4.8327 & 2.8525 & 0.6678 & 0.9967 & [-2.15, 0.37] \\
    0.89  & -1.717 & -1.064 & -0.064 & 0.902 & 4.230 & 2.0561 & 4.6956 & 2.7203 & 0.6314 & 0.9969 & [-2.15, 0.37] \\
    0.90  & -1.742 & -1.097 & -0.102 & 0.872 & 4.091 & 1.9900 & 4.5600 & 2.5836 & 0.5929 & 0.9972 & [-2.15, 0.37] \\
    0.91  & -1.766 & -1.129 & -0.139 & 0.848 & 3.957 & 1.9300 & 4.4252 & 2.4407 & 0.5518 & 0.9976 & [-2.15, 0.37] \\
    0.92  & -1.788 & -1.159 & -0.174 & 0.832 & 3.828 & 1.8767 & 4.2902 & 2.2897 & 0.5074 & 0.9980 & [-2.15, 0.37] \\
    0.93  & -1.808 & -1.189 & -0.209 & 0.825 & 3.704 & 1.8307 & 4.1541 & 2.1285 & 0.4590 & 0.9984 & [-2.15, 0.37] \\
    0.94  & -1.827 & -1.218 & -0.242 & 0.828 & 3.584 & 1.7925 & 4.0161 & 1.9552 & 0.4060 & 0.9989 & [-2.15, 0.37] \\
    0.95  & -1.844 & -1.246 & -0.274 & 0.842 & 3.469 & 1.7628 & 3.8750 & 1.7677 & 0.3477 & 0.9994 & [-2.15, 0.37] \\
    0.96  & -1.859 & -1.272 & -0.304 & 0.868 & 3.359 & 1.7422 & 3.7300 & 1.5637 & 0.2833 & 0.9998 & [-2.15, 0.37] \\
    0.97  & -1.873 & -1.299 & -0.333 & $-$   & 3.253 & 1.8965 & 3.3483 & 0.7489 & $-$    & 0.9991 & [-2.15, 0.37] \\
    0.98  & -1.885 & -1.324 & -0.360 & $-$   & 3.151 & 1.8225 & 3.2835 & 0.7315 & $-$    & 0.9992 & [-2.15, 0.37] \\
    0.99  & -1.897 & -1.348 & -0.386 & $-$   & 3.053 & 1.7515 & 3.2213 & 0.7148 & $-$    & 0.9993 & [-2.15, 0.37] \\
    1.00  & -1.907 & -1.372 & -0.410 & $-$   & 2.960 & 1.6835 & 3.1616 & 0.6988 & $-$    & 0.9995 & [-2.15, 0.37] \\
    1.01  & -1.917 & -1.395 & -0.432 & $-$   & 2.870 & 1.6185 & 3.1042 & 0.6834 & $-$    & 0.9996 & [-2.15, 0.37] \\
    1.02  & -1.925 & -1.417 & -0.453 & $-$   & 2.784 & 1.5564 & 3.0488 & 0.6685 & $-$    & 0.9997 & [-2.15, 0.37] \\
    1.03  & -1.934 & -1.438 & -0.472 & $-$   & 2.701 & 1.4972 & 2.9954 & 0.6539 & $-$    & 0.9998 & [-2.15, 0.37] \\
    1.04  & -1.942 & -1.459 & -0.490 & $-$   & 2.622 & 1.4409 & 2.9438 & 0.6396 & $-$    & 0.9999 & [-2.15, 0.37] \\
    1.05  & -1.951 & -1.479 & -0.506 & $-$   & 2.547 & 1.3873 & 2.8937 & 0.6253 & $-$    & 0.9999 & [-2.15, 0.37] \\
    1.06  & -1.960 & -1.498 & -0.520 & $-$   & 2.475 & 1.3366 & 2.8449 & 0.6110 & $-$    & 1      & [-2.15, 0.37] \\
    1.07  & -1.969 & -1.517 & -0.532 & $-$   & 2.406 & 1.2885 & 2.7974 & 0.5965 & $-$    & 1      & [-2.15, 0.37] \\
    1.08  & -1.980 & -1.535 & -0.543 & $-$   & 2.339 & 1.2431 & 2.7508 & 0.5816 & $-$    & 1      & [-2.15, 0.37] \\
    1.09  & -1.993 & -1.552 & -0.553 & $-$   & 2.276 & 1.2008 & 2.7051 & 0.5663 & $-$    & 1      & [-2.15, 0.37] \\
    1.10  & -2.007 & -1.568 & -0.562 & $-$   & 2.216 & 1.1599 & 2.6601 & 0.5503 & $-$    & 1      & [-2.15, 0.37] \\
    1.11  & -2.024 & -1.584 & -0.569 & $-$   & 2.158 & 1.1221 & 2.6155 & 0.5335 & $-$    & 0.9998 & [-2.15, 0.37] \\
    1.12  & -2.044 & -1.599 & -0.576 & $-$   & 2.103 & 1.0865 & 2.5713 & 0.5159 & $-$    & 0.9998 & [-2.15, 0.37] \\
    1.13  & -2.068 & -1.614 & -0.582 & $-$   & 2.050 & 1.0533 & 2.5273 & 0.4971 & $-$    & 0.9997 & [-2.15, 0.37] \\
    1.14  & -2.095 & -1.627 & -0.587 & $-$   & 1.999 & 1.0223 & 2.4834 & 0.4771 & $-$    & 0.9996 & [-2.15, 0.37] \\
    1.15  & -2.127 & -1.640 & -0.593 & $-$   & 1.950 & 0.9933 & 2.4394 & 0.4557 & $-$    & 0.9996 & [-2.15, 0.37] \\
    1.16  & $-$    & -1.653 & -0.598 & $-$   & 1.904 & 1.0018 & 2.3131 & 0.3359 & $-$    & 1      & [-1.46, 0.37] \\
    1.17  & $-$    & -1.664 & -0.604 & $-$   & 1.859 & 0.9761 & 2.2699 & 0.3128 & $-$    & 1      & [-1.46, 0.37] \\
    1.18  & $-$    & -1.675 & -0.611 & $-$   & 1.816 & 0.9505 & 2.2300 & 0.2923 & $-$    & 1      & [-1.46, 0.37] \\
    1.19  & $-$    & -1.686 & -0.619 & $-$   & 1.774 & 0.9247 & 2.1938 & 0.2747 & $-$    & 1      & [-1.46, 0.37] \\
    1.20  & $-$    & -1.695 & -0.629 & $-$   & 1.734 & 0.8983 & 2.1619 & 0.2608 & $-$    & 1      & [-1.46, 0.37] \\
    1.21  & $-$    & -1.704 & -0.642 & $-$   & 1.695 & 0.8708 & 2.1348 & 0.2509 & $-$    & 1      & [-1.46, 0.37] \\
    1.22  & $-$    & -1.712 & -0.657 & $-$   & 1.657 & 0.8418 & 2.1131 & 0.2458 & $-$    & 1      & [-1.46, 0.37] \\
    1.23  & $-$    & -1.720 & -0.675 & $-$   & 1.621 & 0.8108 & 2.0975 & 0.2462 & $-$    & 1      & [-1.46, 0.37] \\
    1.24  & $-$    & -1.727 & -0.698 & $-$   & 1.585 & 0.7773 & 2.0888 & 0.2528 & $-$    & 1      & [-1.46, 0.37] \\
    1.25  & $-$    & -1.733 & -0.726 & $-$   & 1.550 & 0.7407 & 2.0878 & 0.2665 & $-$    & 1      & [-1.46, 0.37] \\
    1.26  & $-$    & -1.738 & -0.759 & $-$   & 1.515 & 0.7005 & 2.0953 & 0.2881 & $-$    & 1      & [-1.46, 0.37] \\
    1.27  & $-$    & -1.743 & -0.798 & $-$   & 1.481 & 0.6559 & 2.1124 & 0.3186 & $-$    & 1      & [-1.46, 0.37] \\
    1.28  & $-$    & -1.747 & -0.844 & $-$   & 1.447 & 0.6064 & 2.1401 & 0.3590 & $-$    & 1      & [-1.46, 0.37] \\
    1.29  & $-$    & -1.751 & -0.899 & $-$   & 1.414 & 0.5512 & 2.1793 & 0.4103 & $-$    & 1      & [-1.46, 0.37] \\
    1.30  & $-$    & -1.754 & -0.963 & $-$   & 1.380 & 0.4897 & 2.2315 & 0.4738 & $-$    & 1      & [-1.46, 0.37] \\
    \hline
 
\end{tabular}
}
  \label{tab:addlabel}
\end{table*}

\subsection{Application of the Method}

We applied the method to five clusters with different metallicities, i.e. M15, M53, M5, NGC 5466, and 
NGC 7006, as explained in the following. The reason of choosing clusters instead of individual field 
giants is that clusters provide absolute magnitudes for comparison with the ones estimated by means of 
our method. The distance modulus, colour excess and metallicity of the clusters are given in Table 7, 
whereas the $g$ magnitudes and $g-r$ colours are presented in Table 8 and they are calibrated in Fig. 4. The $g$ and 
$g-r$ data of the clusters are taken from \cite{An08}. Also the colour excesses and the distance moduli 
of all clusters and the metallicities of M15 and M5 are taken from \cite{An08}. Whereas the metallicities of 
three clusters, M53, NGC 5466, and NGC 7006, are taken from the authors cited in Table 7. \cite{An08} claimed 
$\lbrack Fe/H \rbrack=-1.99$ and -1.48 dex for the clusters M53 and NGC 7006, respectively. Whereas the 
metallicities in \cite{Santos04}, i.e. $\lbrack Fe/H \rbrack=-1.88$ and -1.35 dex,  provide more accurate 
absolute magnitudes. The metallicities cited by \cite{An08} and \cite{Rosenberg99} for the cluster NGC 5466 
are $\lbrack Fe/H \rbrack=-2.22$ and -2.13$\pm$0.36 dex, respectively. Here again, the metallicity in 
\cite{Rosenberg99} plus its error, i.e. $\lbrack Fe/H \rbrack=-1.17$ dex, provides more accurate absolute 
magnitudes.

\begin{table}
\setlength{\tabcolsep}{2pt}
\begin{center}
\scriptsize{
  \caption{Data for the clusters used for the application of the method.}
    \begin{tabular}{lcccc}
    \hline
    Cluster & $E(B-V)$ & $(g-M_{g})_{0}$ & $\lbrack Fe/H \rbrack$ & Ref.\\
    \hline
    M15     & 0.10  & 15.25 & -2.42 & 1 \\
    M53     & 0.02  & 16.25 & -1.88 & 2 \\
    M5      & 0.03  & 14.42 & -1.26 & 1 \\
    NGC5466 & 0.00  & 16.00 & -1.77 & 3 \\
    NGC7006 & 0.05  & 18.09 & -1.35 & 2 \\
    \hline
    \end{tabular}\\
}
(1) \cite{An08}, (2) \cite{Santos04}, (3) \cite{Rosenberg99}.\\
\end{center}
  \end{table}

We evaluated the $M_{g}$ absolute magnitude by means of the the Eq. (4) for a set of $(g-r)_{0}$ 
colour indices where the clusters are defined. The results are presented in Table 9. The columns give: 
(1) $(g-r)_{0}$ colour index, (2) $(M_{g})_{cl}$, absolute magnitude for a cluster estimated by its 
colour magnitude diagram, (3) $(M_{g})_{ev}$, the absolute magnitude estimated by the procedure (4) 
$\Delta M$, absolute magnitude residuals. Also, the metallicity for each cluster is indicated near 
the name of the cluster. The differences between the absolute magnitudes estimated by the procedure presented 
in this study and the ones evaluated via the colour magnitudes of the clusters (the residuals) lie 
between -0.28 and +0.43 mag. However, the range of 94$\%$ of the absolute magnitude residuals is 
shorter, i.e. $–0.1<M_{g}\leq 0.4$ mag. The mean and the standard deviation of the residuals are 
$<\Delta M>=0.169$ and $\sigma=0.140$ mag, respectively. The distribution of the residuals are 
given in Table 10 and Fig. 4.   

The absolute magnitudes on the RGB at a given colour and metallicity do not change linearly or 
quadratically with age. Instead, the absolute magnitudes gets rapidly fainter for young (and massive) 
stars with a certain $g-r$ and $[Fe/H]$, but shows virtually the same absolute magnitude for all old 
stars, i.e. $t>6$ Gyrs. That is, the gradient of the absolute magnitude respect to a given colour 
and metallicity is greater for a young star relative to an old one.

\begin{figure}
\begin{center}
\includegraphics[scale=0.40, angle=0]{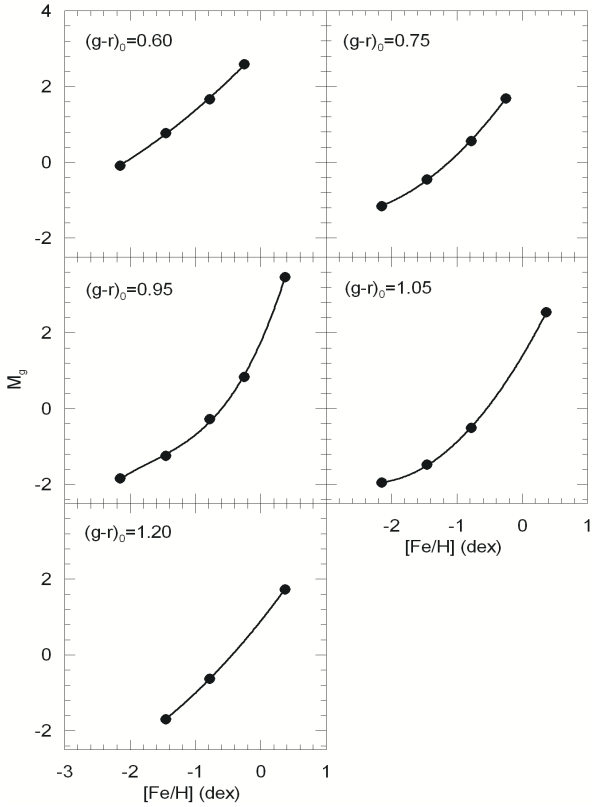} 
\caption[] {Calibration of the absolute magnitude $M_{g}$ as a function of metallicity $[Fe/H]$ for five colour-indices.} 
\label{his:col}
\end{center}
\end {figure}

\begin{figure}[h]
\begin{center}
\includegraphics[scale=0.35, angle=0]{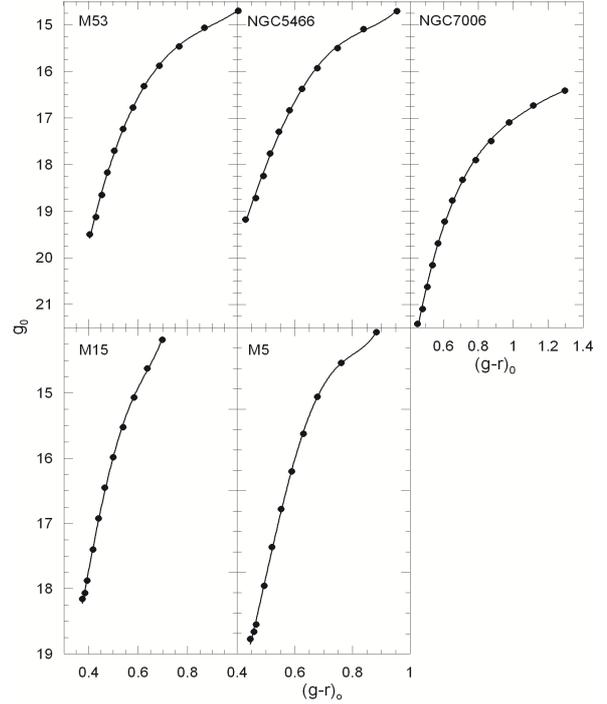}
\caption[] {$g_{0}, (g-r)_{0}$ colour-apparent magnitude diagrams for the Galactic clusters used for the application of the procedure.} 
\end{center}
\end{figure}  

\begin{figure}[h]
\begin{center}
\includegraphics[scale=0.40, angle=0]{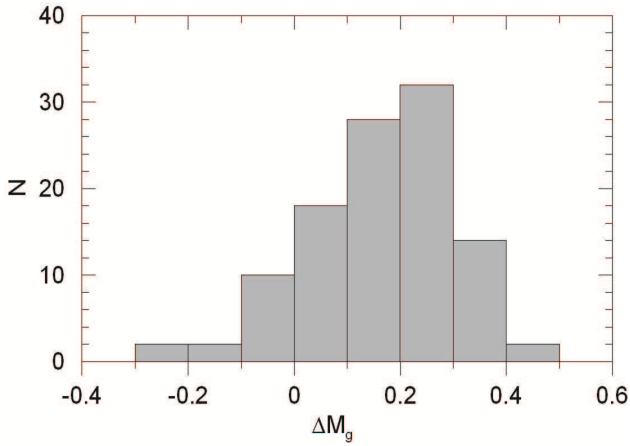}
\caption{Histogram of the residuals.}
\label{histogram}
\end{center}
\end{figure}  

\begin{table*}
  \centering
\tiny{
  \caption{Fiducial giant sequences for the Galactic clusters used in the application of the procedure.}
    \begin{tabular}{cccc|cccc}
    \hline
    $g-r$ & $g$    & $(g-r)_{0}$ & $g_{0}$  & $g-r$   & $g$& $(g-r)_{0}$ & $g_{0}$\\
    \hline
    \multicolumn{4}{c|}{M15}      & \multicolumn{4}{c}{M5 (cont.)} \\
    \hline
    0.803 & 14.553 & 0.697 & 14.181 & 0.552 & 16.802 & 0.520 & 16.690 \\
    0.743 & 14.993 & 0.637 & 14.621 & 0.525 & 17.275 & 0.493 & 17.163 \\
    0.689 & 15.439 & 0.583 & 15.067 & 0.498 & 17.748 & 0.466 & 17.636 \\
    0.645 & 15.895 & 0.539 & 15.523 & 0.490 & 17.840 & 0.458 & 17.728 \\
    0.604 & 16.354 & 0.498 & 15.982 & 0.477 & 17.927 & 0.445 & 17.815 \\
\cline{5-8}    0.571 & 16.821 & 0.465 & 16.449 & \multicolumn{4}{c}{NGC5466} \\
\cline{5-8}    0.546 & 17.296 & 0.440 & 16.924 & 0.954 & 14.704 & 0.954 & 14.704 \\
    0.523 & 17.773 & 0.417 & 17.401 & 0.839 & 15.089 & 0.839 & 15.089 \\
    0.500 & 18.250 & 0.394 & 17.878 & 0.748 & 15.498 & 0.748 & 15.498 \\
    0.490 & 18.440 & 0.384 & 18.068 & 0.679 & 15.929 & 0.679 & 15.929 \\
    0.480 & 18.530 & 0.374 & 18.158 & 0.625 & 16.375 & 0.625 & 16.375 \\
\cline{1-4} \multicolumn{4}{c|}{M53}      & 0.582 & 16.832 & 0.582 & 16.832 \\
\cline{1-4} 1.026 & 14.776 & 1.005 & 14.702 & 0.545 & 17.295 & 0.545 & 17.295\\
    0.890 & 15.140 & 0.869 & 15.066 & 0.514 & 17.764 & 0.514 & 17.764 \\
    0.788 & 15.538 & 0.767 & 15.464 & 0.491 & 18.241 & 0.491 & 18.241 \\
    0.708 & 15.958 & 0.687 & 15.884 & 0.465 & 18.715 & 0.465 & 18.715 \\
    0.645 & 16.395 & 0.624 & 16.321 & 0.429 & 19.179 & 0.429 & 19.179 \\
\cline{5-8}    0.600 & 16.850 & 0.579 & 16.776 & \multicolumn{4}{c}{NGC7006}  \\
\cline{5-8}    0.560 & 17.310 & 0.539 & 17.236 & 1.348 & 16.598 & 1.295 & 16.412\\
    0.525 & 17.775 & 0.504 & 17.701 & 1.168 & 16.918 & 1.115 & 16.732 \\
    0.497 & 18.247 & 0.476 & 18.173 & 1.030 & 17.280 & 0.977 & 17.094 \\
    0.475 & 18.725 & 0.454 & 18.651 & 0.927 & 17.677 & 0.874 & 17.491 \\
    0.451 & 19.201 & 0.430 & 19.127 & 0.839 & 18.089 & 0.786 & 17.903 \\
    0.425 & 19.575 & 0.404 & 19.501 & 0.763 & 18.513 & 0.710 & 18.327 \\
\cline{1-4}     \multicolumn{4}{c|}{M5}       & 0.703 & 18.953 & 0.650 & 18.767 \\
\cline{1-4}    0.915 & 14.165 & 0.883 & 14.053 & 0.659 & 19.409 & 0.606 & 19.223\\
    0.792 & 14.542 & 0.760 & 14.430 & 0.622 & 19.872 & 0.569 & 19.686 \\
    0.710 & 14.960 & 0.678 & 14.848 & 0.591 & 20.341 & 0.538 & 20.155 \\
    0.661 & 15.411 & 0.629 & 15.299 & 0.562 & 20.812 & 0.509 & 20.626 \\
    0.621 & 15.871 & 0.589 & 15.759 & 0.534 & 21.284 & 0.481 & 21.098 \\
    0.584 & 16.334 & 0.552 & 16.222 & 0.504 & 21.604 & 0.451 & 21.418 \\
    \hline
     \end{tabular}
}
  \label{tab:addlabel}
\end{table*}

\begin{table*}
\setlength{\tabcolsep}{2pt}
\center
\tiny{
\caption{Absolute magnitudes ($(M_{g})_{ev}$) and residuals ($\Delta M$) estimated by the procedure explained in our work. $(M_{g})_{cl}$ denotes the absolute magnitude evaluated by means of the colour-magnitude diagram of the cluster.}
        \begin{tabular}{cccc|cccc|cccc}
    \hline
    (1)   & (2)   & (3)   & (4)   & (1)   & (2)   & (3)   & (4)   & (1)   & (2)   & (3)   & (4) \\
    \hline
    $(g-r)_{0}$ & $(M_{g})_{cl}$ & $(M_{g})_{ev}$ & $\Delta M$ & $(g-r)_{0}$ & $(M_{g})_{cl}$ & $(M_{g})_{ev}$ & $\Delta M$ & $(g-r)_{0}$ & $(M_{g})_{cl}$ & $(M_{g})_{ev}$ & $\Delta M$ \\
    \hline

    \multicolumn{4}{c|}{M15 ([Fe/H]=-2.42 dex)} & \multicolumn{4}{c|}{M5 (cont.)} & \multicolumn{4}{c}{NGC7006 ([Fe/H]=-1.35 dex)} \\
    \hline
    0.45  &  1.517 &  1.738 & -0.221 & 0.50  & 2.627 & 2.572  & 0.054 & 0.45  & 3.470 & 3.472 & -0.002 \\
    0.47  &  1.176 &  1.295 & -0.118 & 0.52  & 2.315 & 2.184  & 0.132 & 0.47  & 3.102 & 3.012 & 0.090 \\
    0.50  &  0.719 &  0.697 &  0.021 & 0.55  & 1.863 & 1.705  & 0.158 & 0.50  & 2.596 & 2.427 & 0.169 \\
    0.52  &  0.452 &  0.436 &  0.016 & 0.57  & 1.578 & 1.400  & 0.178 & 0.52  & 2.287 & 2.048 & 0.239 \\
    0.55  &  0.109 &  0.038 &  0.071 & 0.60  & 1.188 & 1.007  & 0.181 & 0.55  & 1.864 & 1.584 & 0.280 \\
    0.57  & -0.085 & -0.122 &  0.037 & 0.62  & 0.955 & 0.781  & 0.174 & 0.57  & 1.606 & 1.279 & 0.327 \\
    0.60  & -0.334 & -0.345 &  0.011 & 0.65  & 0.654 & 0.488  & 0.165 & 0.60  & 1.253 & 0.885 & 0.369 \\
    0.62  & -0.481 & -0.486 &  0.005 & 0.67  & 0.485 & 0.319  & 0.166 & 0.62  & 1.039 & 0.658 & 0.381 \\
    0.65  & -0.692 & -0.688 & -0.004 & 0.70  & 0.279 & 0.097  & 0.182 & 0.65  & 0.747 & 0.364 & 0.383 \\
    0.67  & -0.842 & -0.818 & -0.023 & 0.72  & 0.171 & -0.033 & 0.204 & 0.67  & 0.570 & 0.194 & 0.376 \\
    0.70  & -1.109 & -1.007 & -0.102 & 0.75  & 0.047 & -0.206 & 0.252 & 0.70  & 0.329 & -0.029& 0.358 \\
\cline{1-4}    \multicolumn{4}{c|}{M53 ([Fe/H]=-1.88 dex)} & 0.77  & -0.017 & -0.309 & 0.292 & 0.72  & 0.183 & -0.159 & 0.342\\
\cline{1-4}    0.45  & 2.400 & 2.475 & -0.075 & 0.80  & -0.094 & -0.449 & 0.355 & 0.75  & -0.016 & -0.332 & 0.316 \\
    0.47  & 2.043 & 2.083 & -0.039 & 0.82  & -0.143 & -0.535 & 0.392 & 0.77  & -0.136 & -0.434 & 0.298 \\
    0.50  & 1.557 & 1.570 & -0.014 & 0.85  & -0.230 & -0.656 & 0.426 & 0.80  & -0.301 & -0.573 & 0.273 \\
    0.52  & 1.263 & 1.250 &  0.013 & 0.87  & -0.308 & -0.732 & 0.423 & 0.82  & -0.401 & -0.658 & 0.257 \\
\cline{5-8}   0.55  & 0.866 & 0.842 & 0.023 & \multicolumn{4}{c|}{NGC5466 ([Fe/H]=-1.77 dex)} & 0.85  & -0.538 & -0.723 & 0.185 \\
\cline{5-8}   0.57  & 0.628 & 0.576 & 0.052 & 0.45  & 2.872 & 2.660 & 0.212 & 0.87  & -0.621 & -0.802 & 0.181 \\
    0.60  &  0.309 &  0.225 & 0.085 & 0.47  & 2.526  & 2.263  & 0.263 & 0.90  & -0.737 & -0.916 & 0.179 \\
    0.62  &  0.120 &  0.019 & 0.102 & 0.50  & 2.031  & 1.748  & 0.283 & 0.92  & -0.808 & -0.990 & 0.183 \\
    0.65  & -0.131 & -0.255 & 0.124 & 0.52  & 1.720  & 1.415  & 0.305 & 0.95  & -0.907 & -1.102 & 0.195 \\
    0.67  & -0.279 & -0.417 & 0.138 & 0.55  & 1.284  & 1.000  & 0.284 & 0.97  & -0.968 & -1.259 & 0.290 \\
    0.70  & -0.474 & -0.633 & 0.160 & 0.57  & 1.016  & 0.720  & 0.296 & 1.00  & -1.055 & -1.311 & 0.256 \\
    0.72  & -0.587 & -0.762 & 0.174 & 0.60  & 0.649  & 0.354  & 0.296 & 1.02  & -1.109 & -1.341 & 0.232 \\
    0.75  & -0.737 & -0.934 & 0.197 & 0.62  & 0.429  & 0.140  & 0.290 & 1.05  & -1.187 & -1.380 & 0.193 \\
    0.77  & -0.825 & -1.037 & 0.212 & 0.65  & 0.136  & -0.142 & 0.279 & 1.07  & -1.236 & -1.401 & 0.165 \\
    0.80  & -0.941 & -1.176 & 0.235 & 0.67  & -0.035 & -0.308 & 0.273 & 1.10  & -1.306 & -1.428 & 0.122 \\
    0.82  & -1.010 & -1.259 & 0.248 & 0.70  & -0.257 & -0.527 & 0.270 & 1.12  & -1.351 & -1.445 & 0.093 \\
    0.85  & -1.104 & -1.189 & 0.085 & 0.72  & -0.384 & -0.656 & 0.272 & 1.15  & -1.416 & -1.469 & 0.053 \\
    0.87  & -1.162 & -1.273 & 0.111 & 0.75  & -0.545 & -0.829 & 0.284 & 1.17  & -1.458 & -1.518 & 0.060 \\
    0.90  & -1.245 & -1.391 & 0.146 & 0.77  & -0.635 & -0.931 & 0.296 & 1.20  & -1.518 & -1.545 & 0.027 \\
    0.92  & -1.299 & -1.468 & 0.168 & 0.80  & -0.750 & -1.069 & 0.318 & 1.22  & -1.556 & -1.563 & 0.007 \\
    0.95  & -1.383 & -1.585 & 0.202 & 0.82  & -0.817 & -1.151 & 0.334 & 1.25  & -1.611 & -1.592 & -0.019 \\
    0.97  & -1.441 & -1.751 & 0.310 & 0.85  & -0.909 & -1.064 & 0.155 & 1.27  & -1.645 & -1.615 & -0.030 \\
    1.00  & -1.537 & -1.790 & 0.253 & 0.87  & -0.969 & -1.151 & 0.181 & 1.30  & -1.693 & -1.659 & -0.033 \\
\cline{1-4}\multicolumn{4}{c|}{M5 ([Fe/H]=-1.26 dex)} & 0.90  & -1.067 & -1.275 & 0.208 & $-$    & $-$    & $-$    & $-$ \\
\cline{1-4}            0.45  & 3.392 & 3.668 & -0.276 & 0.92  & -1.143 & -1.357 & 0.214 & $-$    & $-$    & $-$    & $-$ \\
                       0.47  & 3.093 & 3.185 & -0.092 & 0.95  & -1.282 & -1.486 & 0.204 & $-$    & $-$    & $-$    & $-$ \\
    \hline
\end{tabular}
}
 \label{tab:addlabel}
\end{table*}

\begin{table}
  \centering
  \caption{Distribution of the residuals. $N$ denotes the number of stars.}
    \begin{tabular}{rcc}
    \hline
    $\Delta M$-interval & $<\Delta M>$ & N \\
    \hline
    (-0.3, -0.2] & -0.249 &  2 \\
    (-0.2, -0.1] & -0.110 &  2 \\
     (-0.1, 0.0] & -0.033 & 10 \\
      (0.0, 0.1] &  0.045 & 18 \\
      (0.1, 0.2] &  0.163 & 28 \\
      (0.2, 0.3] &  0.258 & 32 \\
      (0.3, 0.4] &  0.348 & 14 \\
      (0.4, 0.5] &  0.425 &  2 \\
    \hline
    \end{tabular}
  \label{residuals}
\end{table}

\begin{table*}
  \centering
\tiny{
 \caption{Absolute magnitudes estimated by altering the metallicity as $[Fe/H]+\Delta[Fe/H]$. The numerical values of $[Fe/H]$ are indicated in the last column. The absolute magnitudes in column (1) are the original ones taken from Table 9, whereas those in the columns $(2)-(5)$ correspond to the increments 0.05, 0.10, 0.15, and 0.20 dex. The differences between the original absolute magnitudes and those evaluated by means of the metallicity increments are given in columns (6)-(9).}

    \begin{tabular}{cccccc|ccccl}
    \hline
          & \multicolumn{5}{c|}{$M_{g}$}               & \multicolumn{4}{c}{$\Delta M$}        \\
    \hline
    {$(g-r)_{0}$} & (1)   & (2)   & (3)   & (4)   & (5)   & (6)   & (7)   & (8)   & (9)   & \multicolumn{1}{c}{}\\
    \hline
    0.50  &  1.570 &  1.651 &  1.732 &  1.813 &  1.893 & 0.081 & 0.162 & 0.242 & 0.323 & $[Fe/H]=-1.88+\Delta[Fe/H]$  \\
    0.65  & -0.255 & -0.205 & -0.153 & -0.099 & -0.044 & 0.050 & 0.102 & 0.156 & 0.211 &  \\
    0.85  & -1.189 & -1.130 & -1.075 & -1.024 & -0.977 & 0.060 & 0.115 & 0.165 & 0.212 &  \\
   \hline
    0.50  &  2.572 &  2.653 &  2.734 &  2.815 &  2.895 & 0.081 & 0.162 & 0.242 & 0.323 &  $[Fe/H]=-1.26+\Delta[Fe/H]$ \\
    0.65  &  0.488 &  0.560 &  0.633 &  0.708 &  0.784 & 0.071 & 0.145 & 0.219 & 0.296 &  \\
    0.85  & -0.656 & -0.616 & -0.575 & -0.531 & -0.483 & 0.039 & 0.081 & 0.125 & 0.173 &  \\
   \hline
    0.50  &  1.748 &  1.829 &  1.910 &  1.990 &  2.071 & 0.081 & 0.162 & 0.242 & 0.323 &  $[Fe/H]=-1.77+\Delta[Fe/H]$ \\
    0.65  & -0.142 & -0.088 & -0.033 &  0.025 &  0.084 & 0.054 & 0.110 & 0.167 & 0.226 &  \\
    0.85  & -1.064 & -1.014 & -0.968 & -0.926 & -0.885 & 0.050 & 0.096 & 0.139 & 0.179 &  \\
   \hline
    0.50  &  2.427 &  2.507 &  2.588 &  2.669 &  2.750 & 0.081 & 0.162 & 0.242 & 0.323 &  $[Fe/H]=-1.35+\Delta[Fe/H]$\\
    0.65  &  0.364 &  0.432 &  0.502 &  0.574 &  0.648 & 0.068 & 0.138 & 0.210 & 0.284 &  \\
    0.85  & -0.723 & -0.686 & -0.648 & -0.608 & -0.566 & 0.037 & 0.075 & 0.114 & 0.156 &  \\
    1.00  & -1.311 & -1.246 & -1.177 & -1.104 & -1.028 & 0.065 & 0.134 & 0.207 & 0.283 &  \\
    1.20  & -1.545 & -1.471 & -1.397 & -1.320 & -1.243 & 0.074 & 0.148 & 0.225 & 0.302 &  \\
    \hline
    \end{tabular}
}
  \label{tab:addlabel}
\end{table*}       

\section{Summary and Discussion}
We presented an absolute magnitude calibration for giants based on the colour-magnitude diagrams 
of six Galactic clusters with different metallicities, i.e. M92, M13, M3, M71, NGC 6791 and NGC 2158. 
All the clusters were observed in the $u'g'r'i'z'$ passbands by \cite{Clem08} except the cluster 
NGC 2158 which is observed in the $ugriz$ passbands by \cite{Smolinski11}. We used the transformations 
of \cite{Rider04} and transformed the $g'$ and $g'-r'$ data in \cite{Clem08} to the $g$ and $g-r$ 
data. Thus, we obtained a homogeneous set of data in SDSS system for absolute magnitude calibration. 
We combined the calibrations between $g_{0}$ and $(g-r)_{0}$ for each cluster with their true distance 
modulus and evaluated a set of absolute magnitudes for the $(g-r)_{0}$ range of each clusters. Then, 
we fitted the $M_{g}$ absolute magnitudes in terms of iron metallicity, $[Fe/H]$, by different degrees 
of polynomials for a given $(g-r)_{0}$ colour index. Our absolute magnitude calibrations cover the range 
0.45 $\leq(g-r)_{0} \leq$ 1.30. However, not all the clusters could be considered for each $(g-r)_{0}$ 
colour index in this interval due to different $(g-r)_{0}$ domains of the clusters. The limited interval 
that all the clusters were considered is 0.85 $\leq(g-r)_{0} \leq$ 0.96. Also, this interval is the unique 
interval where the highest degree (n$ =$ 3) of polynomial was fitted. A linear or quadratic polynomial was 
sufficient for the colour intervals 0.45$\leq(g-r)_{0} \leq$0.84 and 0.97$\leq(g-r)_{0} \leq$ 1.30 for 
a high correlation coefficient.  

We applied the procedure to another set of Galactic cluster, i.e. M15, M53, M5, NGC 5466 and NGC 7006. 
The reason for this chose is that a cluster provides absolute magnitude for comparison with the ones estimated 
by means of our procedure. We used the equations of \cite{Fan99} for de-reddening of the colour and magnitudes, 
and the calibration in Eq. (4) for evaluation a set of $M_{g}$ absolute magnitudes for each cluster in their 
$(g-r)_{0}$ domain.

We compared the absolute magnitudes estimated by this procedure with those evaluated via combination of the fiducial 
$g_{0}$, $(g-r)_{0}$ sequence and the true distance modulus for each cluster. The residuals lie between -0.28 and 
+0.43 mag. However, the range of 94\% of them is smaller, i.e. $–0.1<M_{g} \leq0.4$ mag. The mean and the 
standard deviation of all the residuals are $<\Delta M>=0.169$ and $\sigma=0.140$ mag, respectively. 
The range of the residuals in Paper I was greater than the one in this study, i.e. $-0.61<\Delta M_{V}<+0.66$ mag. 
Also, the mean and the standard deviation of the residuals in a smaller range, -0.4 $\leq \Delta M  \leq$ +0.4, which 
consists of 91\% of the residuals were $<\Delta M>=0.05$ and $\sigma=0.19$ mag, respectively. Comparison of the 
statistical results presented in two studies shows that there is a small improvement on the results in this study 
with respect to the former one. As claimed in Paper I, there was an improvement on the results therein respect to 
the ones of \cite{Hog98}. Hence, the same improvement holds for this study. The same improvement holds also for 
the work of \cite{Ljunggren66}. 

Although age plays an important role in the trend of the fiducial sequence of the RGB, we have not used it as a parameter 
in the calibration of absolute magnitude. Another problem may originate from the red clump (RC) stars. These stars lie 
very close to the RGB but they present a completely different group of stars. Table 10 and Fig. 5 summarize how reliable 
are our absolute magnitudes. If age and possibly the mix with RC stars would affect our results this should show up, 
i.e. the range of the residuals would be greater and their distributions would be multimodal. Whereas, in our study their 
range is small and the histogram of the residuals in Fig. 5 is almost symmetric resembling  a Gaussian distributio. 
Additionally, we should add that the fiducial sequences used in our study were properly selected as RGB. However, 
the researchers should identify and exclude the RC stars when they apply our calibrations to the field stars.  

The accuracy of the estimated absolute magnitudes depends mainly on the accuracy of the metallicity. We altered 
the metallicity by $[Fe/H]+\Delta[Fe/H]$ in the evaluation of the absolute magnitudes by the procedure presented 
in our study and checked its effect on the absolute magnitude. We adopted $[Fe/H]=-1.88$, -1.26, -1.77, -1.35 dex 
and $\Delta[Fe/H]=0.05$, 0.10, 0.15, 0.20 dex and re-evaluated the absolute magnitudes for 14 $(g-r)_{0}$ 
colour indices for this purpose. The differences between the absolute magnitudes evaluated in this way and the corresponding 
ones evaluated without $\Delta[Fe/H]$ increments are given in Table 11. The maximum difference in absolute magnitude is 
$\sim$ 0.3 mag corresponding to the metallicity increment $\Delta[Fe/H]=0.20$ dex. The mean error in metallicity for 
42 globular and 33 open clusters in the catalogue of \cite{Santos04} is $\sigma=0.19$ dex. If we assume the same error 
for the field stars, the probable error in $M_{g}$ magnitudes would be less than 0.3 mag.        

The absolute magnitude could be calibrated as a function of ultraviolet excess, instead of metallicity. However, the 
ultraviolet magnitudes can not be provided easily. Whereas, metallicity can be derived by different methods, such as 
by means of atmospheric model parameters of a star, a procedure which is applied rather extensively in large surveys such 
as RAVE. In such cases, one needs to transform the calibration from SDSS to the system in question. 
The age is a secondary parameter for the old clusters and does not influence much the position of their RGB. 
The youngest cluster in our paper is NGC 2158 with age 2 Gyr \citep{Carraro02}. However, the field stars may be much 
younger. We should remind that the derived relations are applicable to stars older than 2 Gyr. For clarification 
of this argument, let a star younger than 2 Gyr be with colour $0.55<(g-r)_0<0.86$ mag. This star will be more 
metal-rich than the stars in the cluster NGC 2158, and according to the positions of the colour-absolute magnitude 
diagrams of the clusters in Fig. 2, it will be absolutely fainter than a star in the cluster NGC 2158 of the same 
colour. Then, one needs to {\em extrapolate} the corresponding absolute magnitude-metallicity diagram in Fig. 3 for its 
absolute magnitude evaluation (one of the two panels at the top depending on its colour). However, extrapolation 
may result in erroneous absolute $M_g$ magnitudes.   

We conclude that the $M_{g}$ absolute magnitudes of the red giants can be estimated with an accuracy of 
$\Delta M \leq$ 0.3 mag, provided that their $[Fe/H]$ metallicities are known.

\section*{Acknowledgments}
We thank to the anonymous reviewer for his/her comments. This research has made use of NASA's 
Astrophysics Data System and the SIMBAD database, operated at CDS, Strasbourg, France

\end{document}